\documentclass[utf8]{frontiersSCNS} 

\usepackage{url,hyperref,lineno,microtype,subcaption}
\usepackage[onehalfspacing]{setspace}
\usepackage{graphicx}
\usepackage{xcolor}
\usepackage{natbib}
\usepackage{txfonts}
\usepackage{color} 
\usepackage{multirow}
\usepackage{rotating}
\usepackage{lscape}
\usepackage{enumerate}
\usepackage{hyperref}
\hypersetup{colorlinks=true, linkcolor=red, citecolor = blue, filecolor=cyan,  urlcolor=magenta,}

\chardef\us=`\_

\def\keyFont{\fontsize{8}{11}\helveticabold }
\def\firstAuthorLast{Udhwani {et~al.}} 
\def\Authors{Purvi Udhwani$^{1,2}$, Arpit Kumar Shrivastav$^{1,3,4}$, Ritesh Patel\,$^{5}$}
\def\Address{$^{1}$Aryabhatta Research Institute of Observational Sciences, Nainital, UK 263002, India \\
$^{2}$University of Calcutta, 87/1, College Street, Kolkata, WB 700073, India \\
$^{3}$Joint Astronomy Programme and Department of Physics, Indian Institute of Science, Bangalore 560012, India \\
$^{4}$Indian Institute of Astrophysics, Bangalore, India-560034 \\
$^{5}$Southwest Research Institute, 1050 Walnut Street, Suite 300, Boulder, CO 80302, USA}
\def\corrAuthor{Ritesh Patel}

\def\corrEmail{ritesh.patel@swri.org}

\begin{document}
\onecolumn
\firstpage{1}

\title[SITCoM]{SITCoM: SiRGraF Integrated Tool for Coronal dynaMics} 

\author[\firstAuthorLast ]{\Authors} 
\address{} 
\correspondance{} 

\extraAuth{}

\maketitle

\begin{abstract}
SiRGraF Integrated Tool for Coronal dynaMics (SITCoM) is based on Simple Radial Gradient Filter (SiRGraF) used to filter the radial gradient in the white-light coronagraph images and bring out dynamic structures. SITCoM has been developed in Python and integrated with SunPy and can be installed by users with the command \texttt{pip install sitcom}. This enables the user to pass the white-light coronagraph data to the tool and generate radially filtered output with an option to save in various formats as required. 
We have implemented the functionality of tracking the transients such as coronal mass ejections (CMEs), outflows, plasma blobs, etc., using height-time plots and deriving their kinematics. In addition, SITCoM also supports oscillation and waves studies such as for streamer waves. This is done by creating a distance-time plot at a user-defined location (artificial slice) and fitting a sinusoidal function to derive the properties of waves, such as time period, amplitude, and damping time (if any). We provide the provision to manually or automatically select the data points to be used for fitting. SITCoM is a tool to analyze some properties of coronal dynamics quickly. We present an overview of the SITCoM with the applications for deriving coronal dynamics' kinematics and oscillation properties. We discuss the limitations of this tool along with prospects for future improvement.
\section{}

\tiny
 \keyFont{ \section{Keywords:} Python, Software Package, Corona, Coronal Mass Ejection (CME), Solar Oscillations, Coronal Dynamics, Instrumentation and Data Management} 
\end{abstract}

\section{Introduction}
    \label{S-intro}


The white-light solar corona is associated with large-scale eruptions, coronal mass ejections \citep[CMEs;][]{Hundhausen84, Gosling1993JGR, Webb12}, reconnection-based plasmoids \citep[e.g.][]{riley_plasmoids2007ApJ, guo_plasmoids2013ApJ, patel2020A&A, Lee2020ApJ}, coronal inflows \citep[e.g.][]{Wang1999GeoRL, Sheeley2001ApJ, Sheeley2007ApJ, Sheeley2014ApJ, Hess2017ApJ}, streamer waves \citep{Chen2010ApJ, Feng2013ApJ, Decraemer2020ApJ}. All these phenomena are seen to exhibit a wide range of properties and are primarily observed and studied with white-light coronagraph imagery of Large Angle Spectrometric Coronagraph \citep[LASCO;][]{Brueckner95} of Solar and Heliospheric Observatory (SOHO) and Sun Earth Connection Coronal and Heliospheric Imager \citep[SECCHI;][]{Howard2008SSRv} onboard Solar Terrestrial Relations Observatory (STEREO). It has been recently emphasized that most of the dynamics occur in the region from 1.5-6 R$_\odot$ defined as the middle corona \citep{West2022arXiv}. To capture such a wide variety of phenomena, it becomes important to separate these structures from the background and, at the same time, enhance their visibility in the coronagraph FOV of observation.

A couple of algorithms have been developed recently for white-light datasets to separate out the dynamic structures and reduce coronal intensity gradient. 
Normalizing radial gradient filter \citep[NRGF;][]{NRGF2006SoPh} and
Simple Radial Gradient Filter \citep[SiRGraF;][]{Patel2022SoPh, patel2022thesis} are two such processing algorithms. Along similar lines, some algorithms have been used to enhance the off-limb structures in extreme ultraviolet images \citep{Masson2014ApJ, ciisco2020, Seaton2021NatAs, Seaton2023arXiv}. In the manual CME detection catalog of NASA Community Data Analysis Workshop \citep[CDAW;][]{Yashiro04}, CMEs can be tracked in their web interface by manually clicking the points in the running difference images. Automated CME detection has been performed with the aid of feature detection and image processing algorithms such as Computer Aided CME Tracking \citep[CACTus;][]{Robbrecht04, Pant2016}, Solar Eruptive Events Detection System \citep[SEEDS;][]{Olmedo08}, Automatic Recognition of Transient Events and Marseille Inventory from Synoptic maps \citep[ARTEMIS;][]{Boursier09}, Coronal Image Processing \citep[CORIMP;][]{Byrne12, Morgan12}, CMEs Identification in Inner Solar Corona \citep[CIISCO;][]{ciisco2020} and algorithms based on machine learning \citep{MLQu2006, ML2017}. These algorithms have been mainly designed for and applied on large-scale transients, but not on small-scale flows and blobs-like structures.

Streamers are large-scale quasi-static bright structures and are seen to be stretched from the inner to outer corona in the eclipse images \citep{1992Koutchmy}. CMEs are observed to displace streamers sideways and generate streamer oscillations \citep{2010Chen, 2013Feng, 2013Kwon}. The statistical study on streamer waves suggested that they can be a plausible candidate for coronal seismology \citep{2020Decraemer}. In the context of automatic detection of transverse oscillations in the solar corona, Northumbria University Wave Tracking \citep[NUWT;][]{Morton2016zndo} was developed and extended by \citep{Weberg2018ApJ}. The use of NUWT can be advantageous in carrying out statistical analysis of transverse waves. However, the NUWT code is best applicable for decayless waves observed in coronal loops \citep{Weberg2018ApJ}. Furthermore, the automatic identification and fitting of waves in time-distance ($x-t$) diagrams can be affected by the complex and varying backgrounds, as well as the presence of neighboring structures, which may necessitate manual fitting procedures. Additionally, there is a lack of a tool for quickly analyzing wave properties.

In recent times, Jhelioviewer \citep{JHelio}, an open-source data visualization tool was developed with basic image processing and integration of different catalogs for quick analysis before proceeding to an in-depth one. An automated data-mining-based web interface combined with feature detection has been developed as a part of the Heliophysics Events Knowledgebase \citep[HEK;][]{Hurlburt2012SoPh}. Using the multi-instrument observations and tracking CMEs, solar energetic particles (SEPs), co-rotating interaction regions (CIRs), etc a community-led effort by European Space Agency’s Modelling and Data Analysis Working Group (MADAWG) developed several tools, e.g. magnetic connectivity tool \citep{Rouillard2020A&A}. Another web-based interface to track the CMEs, SEPs, and CIRs from the inner corona to the outer heliosphere is the `Propagation Tool' \citep{Rouillard2017P&SS}.
Solar MAgnetic Connection Haus \citep[SolarMACH;][]{Gieseler2023FrASS}, another open-source tool was developed recently for the visualization of magnetic configurations and connections with various spacecraft and planetary bodies in the heliosphere. To track the CMEs based on multi-viewpoint observations from STEREO and SOHO coronagraphs, a web-based application called STEREO CME Analysis Tool \citep[StereoCAT;][]{Millward2013} is hosted at NASA Community Coordinated Modeling Center (\url{https://ccmc.gsfc.nasa.gov/analysis/stereo/}).

All these tools have been supported with basic image processing algorithms and are helpful in quick analysis serving their purposes. Often the output of these tools has also been part of in-depth scientific analysis. In this paper we introduce an open-source tool, SiRGraF Integrated Tool for COronal dynaMics (SITCoM), utilizing the SiRGraF algorithm and bringing out the dynamic features of the solar corona. This tool is currently targeted for studying the kinematics of CMEs, plasmoids, streamer blobs, outflows, inflows, etc., and the analysis of streamer waves and similar oscillatory phenomena. This paper is arranged as follows: in Section \ref{S-design} we introduce and explain the architecture of the tool. The application of SITCoM for two broad aspects with illustrative examples is presented in Section \ref{S-data} followed by a summary and discussions along with the scope of future improvements in Section \ref{S-summary}.

\section{Software Design}
    \label{S-design}
    
\texttt{sitcom} is fully developed using \textsc{Python} and is temporarily made available as a PyPi package\footnote{PyPi: \url{https://pypi.org/project/sitcom/}} and is mainly compiled using PyQt and has other small dependencies. It can be installed using the command \newline \texttt{pip install sitcom}. \\ The software can be run for further use by the command in the terminal\\
\texttt{python3 -m sitcom} \\
Further updates will be released that will make it more flexible for different platforms. It is available on Github\footnote{GitHub:\url{https://github.com/pu3/SITCoM}} as well for a better look at the code and to raise issues if any.
\\

\begin{figure*} 
    \centering 
        \includegraphics[width=1\linewidth]{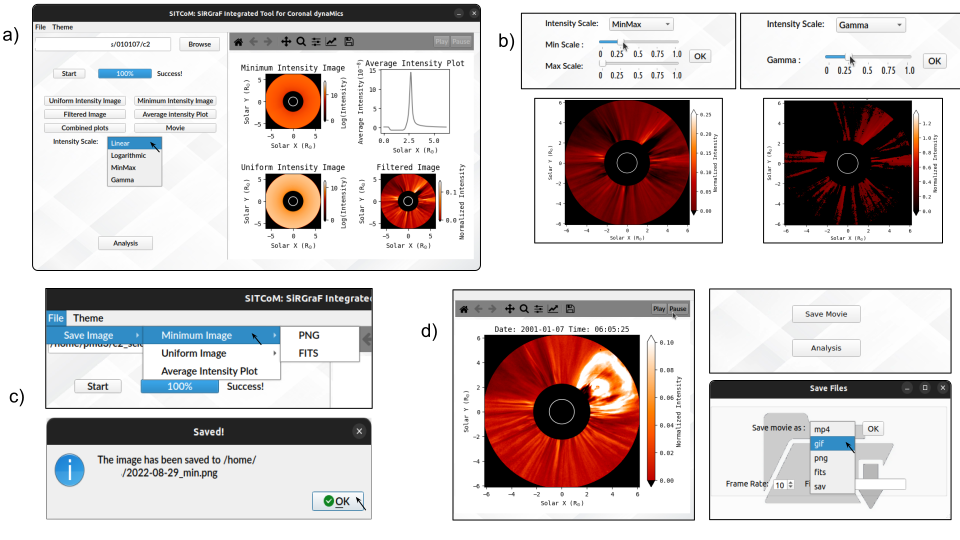}
        \caption{\textbf{a.} An inclusive overview of the tools available in the SiRGraF part of the software including the availability to browse the path or add the path to data. 
        \textbf{b.} The intensity scaling options  and the respective output below as applied on LASCO-C2 images. 
        \textbf{c.}The toolbar menu options to save images and to switch themes. On the bottom is the dialog box that appears when the `Minimum Image' is saved displaying the location of the saved file. 
        \textbf{d.} On the \textit{Left}, is a screenshot of a paused frame of the movie, and on the \textit{Right} below are the available formats in which you may save the movie. The `Save movie' button appears when the `Movie' button is clicked.}
        \label{sirgraf_overview}
\end{figure*}

\subsection{SiRGraF interface}
    \label{ss-interface}


The initial part of this Graphical User Interface (GUI) is shown in Figure \ref{sirgraf_overview} where the first step is the implementation of the SiRGraF algorithm. 
The interface is designed in such a way that the only task of the user is to provide the path containing FITS files in the given address bar by pasting a copied path or use the browse option available. The path for LASCO/C2 images for January\,07, 2001 is shown for illustration.  After providing the path of the data, the start button is used to initialize the process of implementation of SiRGraF and the status of which can be tracked in the status bar. Once the process is completed, the right panel in the GUI shows the minimum intensity, and uniform intensity images along with the average intensity profile and a sample filtered image after pressing the respective buttons as shown in Figure \ref{sirgraf_overview}a. By default, linear intensity scaling is used to display all the generated image outputs.

The images filtered with SiRGraF can be visualized in an embedded matplotlib-based player consisting of basic functionalities like Play, Pause, and Zoom in the toolbar, this, among other options to assess these images, will help in better evaluation of the data. This section also allows the user to perform the basic intensity scaling like logarithmic, min-max, and gamma correction \citep{Gonzalez2002dip}. Figure \ref{sirgraf_overview}b shows some of the different intensity scaling options available and the respective output when applied to SiRGRaF processed LASCO/C2 images. Utilizing these options will help highlight certain features against others which can be used in the subsequent analysis section. 
In conjunction with that, the user can save individual elements of SiRGRaF like the minimum and uniform images from the `File' option of the menu bar (Figure \ref{sirgraf_overview}c). Additionally, the movie can be saved in `.mp4', `.gif' (video formats), `.png', `.fits', and `.sav' (saves each processed frame with the file name with the timestamp from the movie. For example, if the file name is `cme', it will save files with one of the names as `cme-\_MMDDTHHMMDD.png'). Along with this, an option to control the frame rate is also provided to save video formats. It creates a hierarchical structure of folders to save these files. For example, for a data file dated 6th July 2004 it will first create a folder named `2004-07-06' in the `/home/' directory and sub-folders within that `PNG', `FITS', and `SAV' depending on the chosen format to save. The movies will be saved in the `2004-07-06' folder.

An optional tool available in the menubar is to switch the GUI theme to dark as it has lately become a preference of users worldwide. 
Once the feature of interest is identified, the user can proceed to the analysis section by using the `Analysis' button.
\\

\subsection{Analysis Interface}
    \label{ss-analysis}

\begin{figure*} 
    \centering 
        \includegraphics[width=1\linewidth]{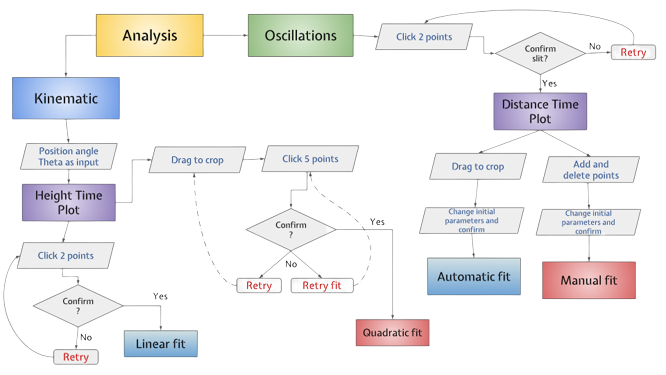}
        \caption{Mechanism of the analysis interface showing the steps required to obtain distance-time plots, and consequently perform respective fits on them}
        \label{analysis_fc}
\end{figure*}

An overview of how the analysis using \texttt{sitcom} is presented through a flow chart in Figure \ref{analysis_fc}.  Upon clicking the `Analysis' button, a user is presented with options to estimate either the kinematic or oscillation properties of the dynamic event. The flow chart illustrates the series of processes that are executed when utilizing the application. 

\begin{figure*} 
    \centering 
        \includegraphics[width=1\linewidth]{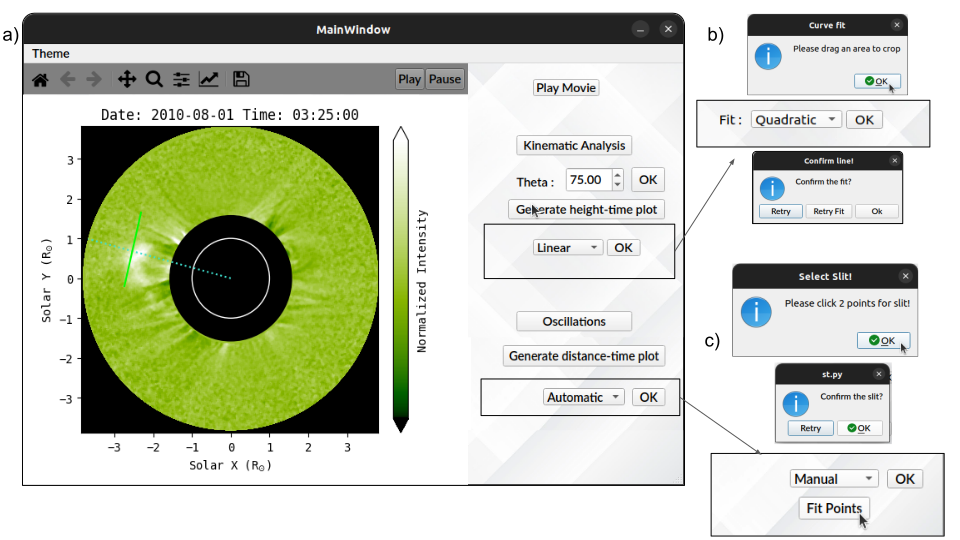}
        \caption{\textbf{a.} An overview of the analysis interface showing the PA with a dotted line and selected slit with a solid green line. 
        \textbf{b.} Dialogue boxes that pop up when performing a quadratic fit \textbf{c.} Option to select slit after the user presses the ``Oscillations" button. The fitting can be performed manually or using automatic fit selection. }
        \label{analysis_overview}
 \end{figure*}

Figure \ref{analysis_overview} provides a snapshot of the GUI corresponding to the analysis interface. Clicking the analysis button takes the user to another window that is adjacent to the main interface. This section can also be accessed directly after the ``Start" button has been clicked i.e. This window contains two provisions that can be used for kinematics analysis and carry out waves or oscillation studies.

The kinematic analysis section allows the user to generate height-time plots at various position angles and perform linear or higher-order fitting on selected data points. The fittings provide information about the velocity and acceleration of the feature, depending on the type of fitting used. The ``Retry" option is available to help obtain better-fit values if required. Additionally, the option for linear fitting is selected by default and can be updated through a drop-down menu. To perform a Linear fit, the user can first use the toolbar zoom and then click two points (Section \ref{ss-CMEkine}). As shown in \ref{analysis_overview}b, for quadratic fitting the user is required to first drag an area to zoom in and then click five points for fitting. Two retry options are available: clicking the ``Retry" button plots the original height-time graph, allowing the user to zoom again on a different region by dragging a rectangle, while clicking ``Retry Fit" clears the currently fitted points, enabling the user to perform the fit again for the already zoomed region. These options can be availed as many times as needed. After the fit is satisfactory, it can be confirmed by `Ok' and will display a plot with velocity and/or acceleration obtained from the fit.

Another option is to analyze oscillations by placing an artificial slit across a feature, which generates a distance-time plot over the length of the slit. The ``Oscillation" tab on the interface allows users to choose an artificial slit of their desired position by clicking two points in the movie frame. Once the slit is selected, a dialogue box will pop up to confirm the selection. If the user wants to re-select the slit, they can use the ``Retry" option on the pop-up box (Figure \ref{analysis_overview}c). After selecting the slit, the user can produce distance-time ($x-t$) maps by clicking on the option ``Generate distance-time plot", with ``Automatic fit" set as the default option. If there are any oscillations in the $x-t$ map, the user can fit these oscillations by selecting the region using drag crop around the oscillation (for automatic fit) or toolbar zoom option (for manual fit), and then perform fit through a series of steps. This fitting procedure is discussed in Section \ref{ss-streamer} with examples.
\\
  
\section{Data Used and Software Functionality}
    \label{S-data}

We acquired the level 0.5 data of the C2 coronagraph of LASCO from the Virtual Solar Observatory (VSO) for 29th August, 2022. We converted the data into Level 1 using the \textit{reduce\_level\_1.pro} procedure available in the solar software (SSW) to correct for the image rotation and calibration. Other LASCO/C2 data presented in this paper were already level 1 corrected. Apart from these, STEREO/COR-1 data has also been used. Level 0.5 polarized brightness triplets (0$^{\circ}$, 120$^{\circ}$, 240$^{\circ}$) of COR-1 are acquired from VSO for the date of 1st August, 2010. These images are reduced to level 1 using the \textit{secchi\_prep.pro} procedure of STEREO pipeline. The triplets are combined to obtain the total brightness images which are also calibrated and rotation corrected in this step. The level 1 data of LASCO/C2 and COR-1 are provided to the SITCoM analysis tool as FITS files.
\\
\subsection{Kinematics of Transients}\label{S-kinematics}
    
\subsubsection{CME case}\label{ss-CMEkine}

\begin{figure*} 
    \centering 
        \includegraphics[width=1\linewidth]{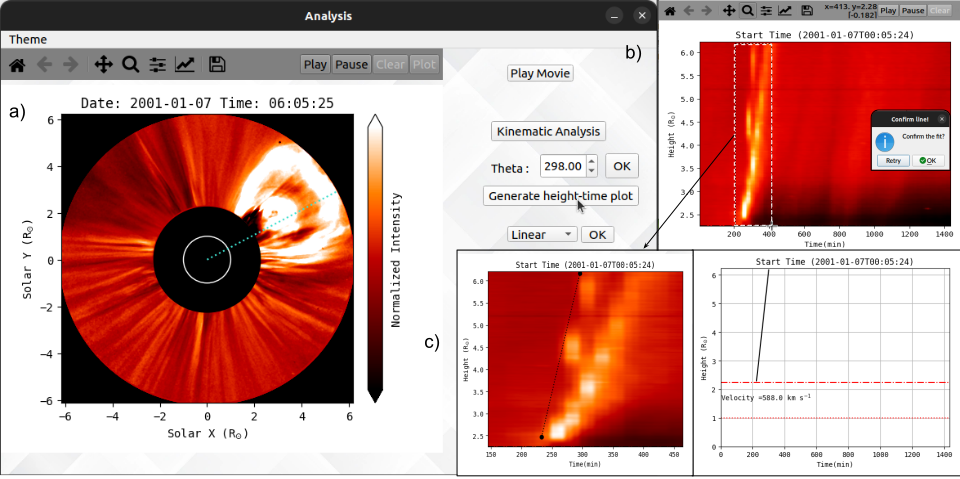}
        \caption{\textbf{a.} Movie frame showing a dashed line at the PA of 298$^{\circ}$ \textbf{b.} Usage of the zoom tool to crop the region for fitting on the height-time plot \textbf{c.} Linear fit confirmed after clicking 2 points showing a velocity of 588\,km\,s$^{-1}$. }
        \label{linear_fit}
\end{figure*}

CMEs are identified as propagating white-light large-scale structures in the coronagraph field of view (FOV) \citep{Hundhausen84}. An application is illustrated in Figure \ref{linear_fit} with LASCO/C2 data-set of January 07, 2001, where a CME occurred at the north-west limb of the Sun. This CME has also been shown as a representative example for demonstrating NRGF by \citet{NRGF2006SoPh} and for SiRGraF by \citet{Patel2022SoPh}. It is known that CMEs after reaching heights beyond $\sim$3\,R$_\odot$ attain a constant speed \citep{Majumdar2020ApJM}. This height also lies in the middle corona within the LASCO field of view and hence CMEs can be considered to have a linear profile in the height-time plot. As shown in Figure \ref{linear_fit}, any position angle can be provided to generate the height-time plot, for this case, we have used a PA of 298$^{\circ}$. Once it is generated, the bright ridge corresponding to the CME can be zoomed in using the toolbar zoom icon, and the leading edge can be fitted with a straight line by clicking two points (after de-selecting the zoom icon). The interface provides an option to retry and finalize the points for the fit, to confirm it. Once confirmed, a plot with the value for the velocity of the CME will be displayed. For the example presented here, the speed estimated here of 588\,km\,s$^{-1}$.  matches well with the value in CDAW catalog as 633\,km\,s$^{-1}$.

\begin{figure*} 
    \centering 
        \includegraphics[width=1\linewidth]{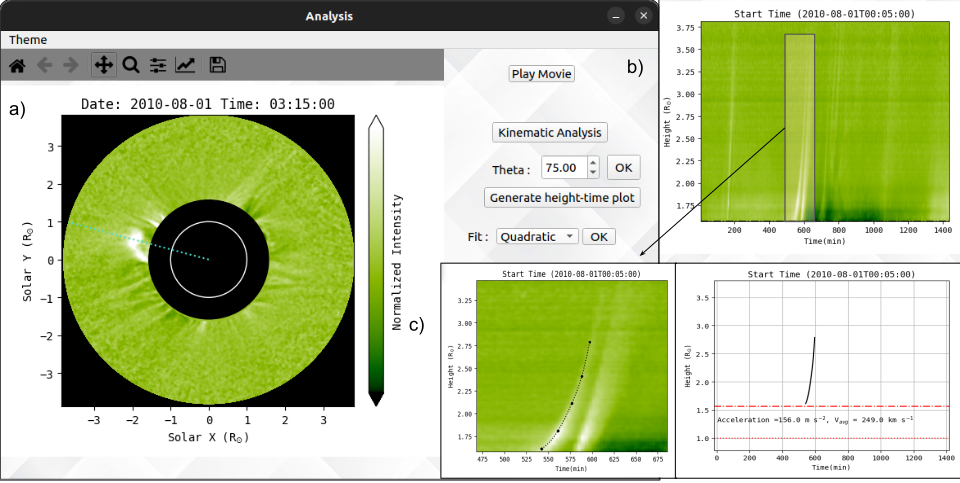}
        \caption{\textbf{a.} Movie frame showing a line at the PA of 75$^{\circ}$ \textbf{b.} Usage of the drag rectangle to crop the region for fitting on the height-time plot \textbf{c.} Quadratic fit confirmed after clicking 5 points showing the values of acceleration and average velocity. }
        \label{quadratic_fit}
\end{figure*}

Most of the CMEs which are observed in STEREO/COR-1 are still undergoing an acceleration phase. This is also seen in an example shown in Figure \ref{quadratic_fit} where a CME is observed in COR-1B on August 01, 2010, at the north-east limb. The height-time plot is generated along the dotted green line shown in the coronagraph image and is shown in the b) panel of the same Figure. The leading edge of the CME is seen with a curved profile suggesting the presence of acceleration. Using the available quadratic fit option, data points are selected in the height-time plot. The points are confirmed for fit by visual inspection. We found that this CME showed an acceleration of $\approx$156\,m\,s$^{-2}$ in the COR-1B FOV, and an average speed of 249\,km\,s$^{-1}$.

\subsubsection{Plasma Blobs case}
    \label{ss-blob}
\begin{figure*} 
    \centering 
        \includegraphics[width=1\linewidth]{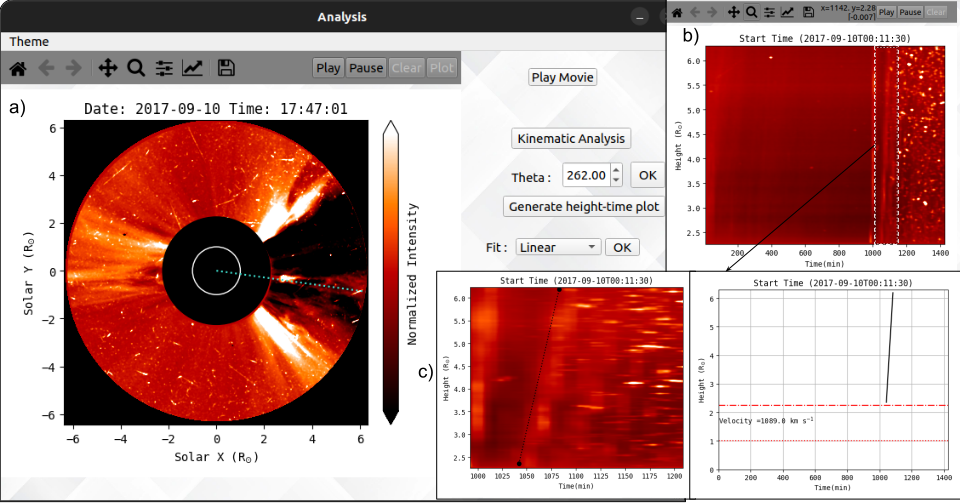}
        \caption{\textbf{a.} Movie frame with the dotted line at the PA of 262$^{\circ}$ along the plasma blob \textbf{b.} Usage of the zoom tool to crop the region for fitting on the height-time plot \textbf{c.} Linear fit on the ridge of plasma blob, showing a velocity of 1089\,km\,s$^{-1}$ }
        \label{plasma_blob}
\end{figure*}

Plasma blobs arising due to magnetic re-connection have been considered as one of the transient structures formed in the Sun's corona. These may be associated with CMEs often observed in coronagraph images post-CME rays \citep{Webb2016SoPh}, and the streamer blobs originating at the tip of streamers. These blobs appear as small-scale discrete white-light structures with nearly elliptical shapes in coronagraph images. 
Figure \ref{plasma_blob} shows one of the many plasma blobs observed along the post-CME current sheet of September 10, 2017. The right panel of the Figure shows the distance-time plot generated for the position angle of 262$^{\circ}$. The first ridge corresponds to the leading edge of the CME followed by plasma blobs. It can be noticed in the distance-time plot that the observed blobs show inclined ridges, hence, we used linear fitting to characterize the speed. The velocity of the blob is displayed after confirming the plot fit. The estimated speed of $\approx$1000\,km\,s$^{-1}$ agrees with the speed distribution of plasma blobs observed in LASCO/C2 FOV for this data set reported by \citet{patel2020A&A}. The distance-time plot also reveals more ridges following the one used for analysis. These correspond to the successive blobs which can be approached in a similar way to measure their speeds.

\subsubsection{Inflow case}
    \label{ss-downflow}

\begin{figure*} 
    \centering 
        \includegraphics[width=1\linewidth]{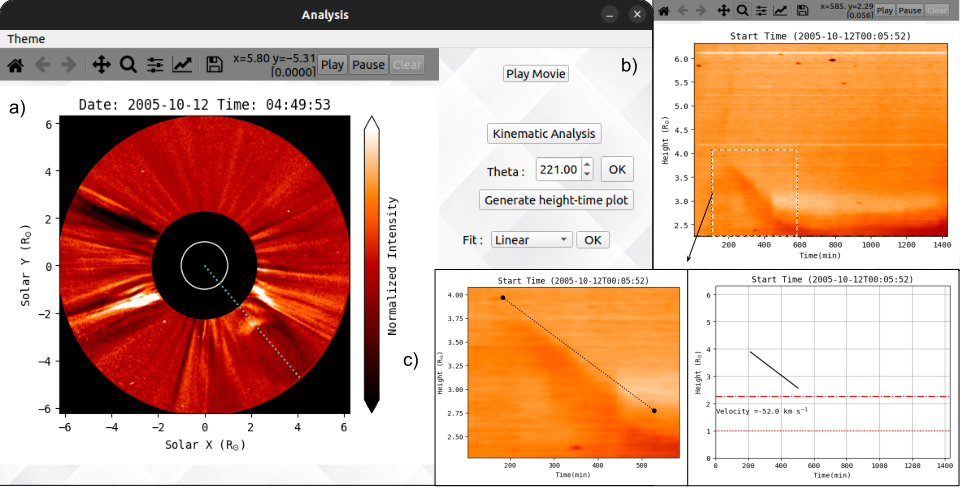}
        \caption{\textbf{a.} The dark curve as seen in the movie frame along the PA of 221$^{\circ}$ is the observed inflow in LASCO FOV \textbf{b.} Usage of the zoom tool to crop the region of the dark ridge  \textbf{c.} Linear fit performed on the inner edge, showing a velocity of -52\,km\,s$^{-1}$, indicating in-flowing material. }
        \label{inflow}
\end{figure*}

Often material has been observed falling toward the Sun after a CME has passed in the LASCO FOV. This in-falling material can be seen as a result of partial eruption or dark inflows which are in the wake of CME. These structures have been tracked manually in many studies in the past. The SITCoM interface has been tested to track such features using LASCO/C2 images. An example of dark inflow is shown in Figure \ref{inflow} where these flows are observed on October 12, 2005, moving towards the Sun after the CME has passed in LASCO FOV. The dotted line represents the position angle of 221$^\circ$ used to track the inflow. The height-time plot is generated corresponding to this dotted line and is shown in Figure \ref{inflow}b. The colormap of this plot is inverted for the visualization of the ridge corresponding to the inflow. This ridge shows straight-line behavior and therefore linear fitting was used to estimate its speed. The fitted curve is shown in panel (c). The negative value of the measured speed indicates that the structure is moving toward the Sun. In case curved ridges are observed for the sunward moving material, as in the studies such as \citet{Tripathi2007A&A, Longcope2018ApJ}, then quadratic fit can be used to measure the kinematics of such flows. \\

\subsection{Oscillations and Waves}
\label{S-oscillations}
\subsubsection{Streamer Wave case}
        \label{ss-streamer}
To demonstrate the analysis of oscillations in helmet streamers perturbed by coronal mass ejections (CMEs), we present two case studies. In this section, we outline the methodology used, involving the SiRGraF algorithm as described in Section \ref{ss-interface}, and highlight the analysis tool employed for accurate data interpretation.
By choosing the ``Play Movie" option in the interface (Figure \ref{analysis_overview}a), users can access a movie displaying the SiRGraF processed images. This feature assists in identifying the optimal placement for the artificial slit, by clicking two points on the frame, corresponding to the desired location. The interface allows for multiple retries, ensuring the selection of the most appropriate slit placement.\\


\subsubsection{Manual fitting} 
    \label{manual}

\begin{figure*} 
    \centering 
        \includegraphics[width=1\linewidth]{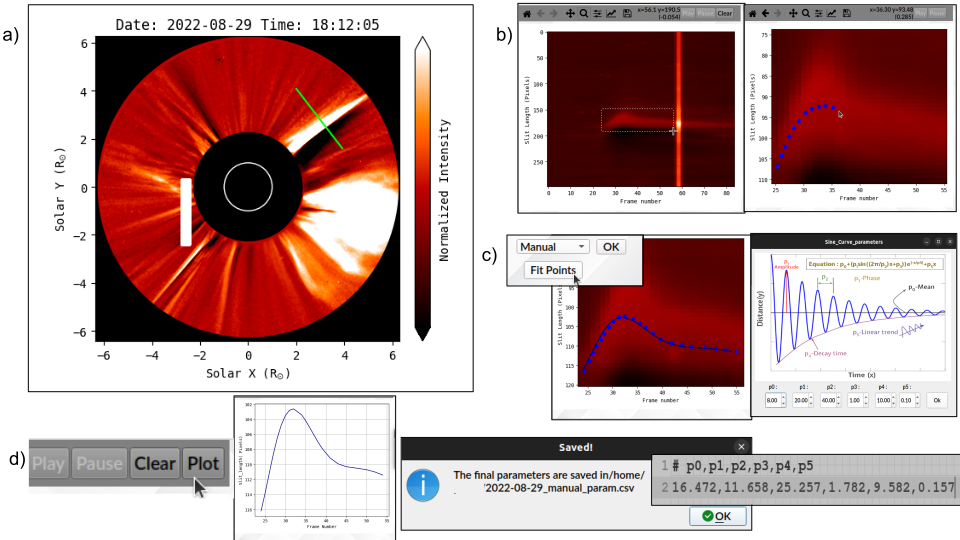}
        \caption{ \textbf{a.} Selected slit on the movie frame from LASCO C2 data of 29 August 2022, perpendicular to the streamer. 
        \textbf{b.} The required area to fit on the generated distance-time plot is first zoomed on and then points are selected for fit. 
        \textbf{c.} ``Fit Points" button performs a fit on the current points based on the initial guess and pops open a window to modify parameters for this fit.
        \textbf{d.} Next, this fit is plotted and parameters saved in a `.csv' file as shown in the dialogue box
        The movie is available in the \hyperref[Movies/OSC_manual_fit.mp4]{online} version}
        \label{manual_fit}
\end{figure*}

We use the set of LASCO/C2 level-1 images of August 29, 2022, between 15:12 and 23:48 UT for the illustration of the manual fitting case, where a CME was observed near the western limb, perturbing the streamer as it propagated in the radial direction. We selected the slit perpendicular to the streamer (Figure \ref{manual_fit}a). The ``Generate distance-time plot"  button on the interface collates the intensities at the slit position for every frame and produces an $x-t$ map. While still under development, the features provided in manual fitting are, to perform fit on the distance-time plot through a series of following simple steps implemented in the current version. 

\begin{enumerate}
        \item Click on the zoom icon and crop the region of interest
        \item Deselect the zoom icon and use the left-click button to add a point where the mouse pointer is. The right-click button deletes the last added point.(Figure \ref{automatic_fit}b)
        \item Once the points are ready for fit, click on the `Fit Points' button (Figure \ref{automatic_fit}c), which will open a window displaying the parameters of the current fit and will fit the current points based on this initial guess. The profile of selected points will be fitted using a decaying sinusoidal function of the following form \citep{Nakariakov1999} :

        \begin{equation}\label{eq1}
                y(x) = p_{0}+p_{1} \sin \left(\frac{2\pi x}{p_{2}}+p_{3} \right) e^{\frac{-x}{p_{4}}}+p_{5}x,
        \end{equation}

        where $p_{0}$ is the mean position, $p_{1}$ is the initial amplitude of the oscillation, $p_{2}$ is the period of oscillation, $p_{3}$ denotes the phase, $p_{4}$ indicates the damping time and $p_{5}$ is the coefficient for the linear trend.
        
        \item These parameters can be updated or changed to modify the fit according to the needs of the user to attain a better fit. 
        \item After that, the `Plot' button can be clicked to generate the final curve used for fit, and their respective parameter will be automatically saved into a `.csv' file in the `/home/' directory.
\end{enumerate}  

The user can click on the `Generate distance-time plot' button as many times as required to retry the fit from the beginning. For this particular case of the streamer, we got the initial perturbation amplitude, period, and damping time as 0.14\,R$_\odot$, 253 minutes, and 96 minutes, respectively, which are consistent with previous streamer wave events depicted in  \cite{2010Chen, 2013Kwon, 2020Decraemer}. The period and damping time relation at different heliocentric distances can be used to check the wave mode of the streamer wave \textbf{\citep{2002Ofman}}. \\

\subsubsection{Automatic fitting} 
    \label{automatic}
\begin{figure*} 
    \centering 
        \includegraphics[width=1\linewidth]{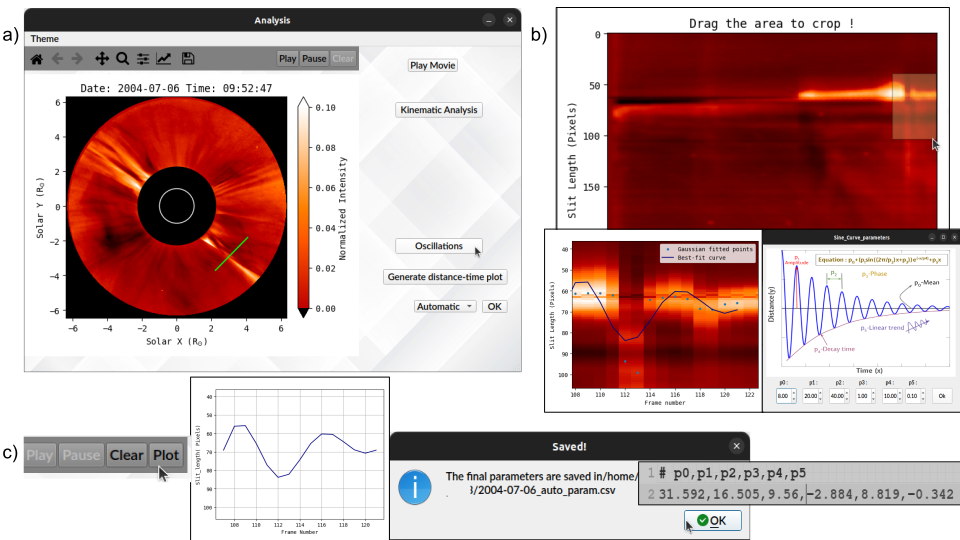}
        \caption{ \textbf{a.} Selected slit on the movie frame from LASCO C2 data of 06 July 2004, perpendicular to the streamer. 
        \textbf{b.} Selected area on the generated distance-time plot, showing oscillations is first zoomed on, and then the automatic fitting is performed with an initial guess.  
        \textbf{c.} After plotting the final fit, the parameters of the fit will be saved in a `.csv' file} 
        \label{automatic_fit}
\end{figure*}

\
We use LASCO/C2 level 1 images of July 06, 2004 between 00:04 and 23:52\,UT for the illustration of the automatic fitting case. 
A streamer is perturbed by a CME located near the south-west region in the LASCO image (see Figure \ref{automatic_fit}). We selected the slit perpendicular to the streamer and used the ``Generate distance-time plot" to produce a $x-t$ map. As shown in the Figure, after the plot is generated, the user can select the constricted part where the oscillation occurs by dragging a rectangle around that region. At every time step (corresponding to columns in $x-t$ map), a Gaussian profile is fitted across the slit length to find out the position of the oscillating streamer. An initial automatic fit, based on equation  \ref{eq1}, will be performed on that region (Note: User may have to try for different rectangles around the oscillation for better fit). This part can be retried many times by clicking on the ``Generate distance-time plot" button. After the fit is performed with the initial guess, another window shows up with these parameters (similar to manual fitting \ref{ss-streamer}), which can be modified to improve the fit. The automatic fit can be advantageous over manual fit as it can remove the human bias created by selecting the points on the x-t map. We found the initial amplitude, period, and damping time from the automatic fit as 0.2\,R$_\odot$, 95 minutes, and 88 minutes, respectively. Using a sample of 22 events, \cite{2020Decraemer} determined the average period of streamer waves as 239 minutes, a similar order of magnitude  values are obtained using manual and automatic fit techniques in this work. The period determination could be further used to calculate phase speed and an estimate of background solar wind speed considering the wavelength is known \citep{2020Decraemer}. The measurement of phase speed could also be useful in determining the magnetic fields in the streamers \citep{2011Chen}. In brief, SITCoM is a useful tool for quickly analyzing the oscillation properties of streamer waves. \\

\section{Summary and Future Developments}
    \label{S-summary}

We introduce a new tool, SITCoM, which is built upon the SiRGRaF algorithm for processing white-light coronagraph images. The application enables rapid analysis of coronal dynamics such as CMEs, plasma blobs, streamer waves, and similar transients, observed in white-light coronagraph data. The output of SITCoM can serve as preliminary analysis for in-depth scientific investigations or can be integrated into larger studies using various datasets.
\texttt{SITCoM} is developed as an open-source application that features a simple user interface based on PyQT and other minor dependencies in Python. It can be installed using pip command in python.

The SITCoM graphical user interface (GUI) incorporates the SiRGRaF algorithm in its initial stage, which proves to be a simple yet effective tool for highlighting dynamic features in the corona. The dynamic features are enhanced against the background throughout the coronagraph FOV. We have illustrated the usage of SITCoM using LASCO and STEREO or white-light coronagraph data for various test cases. Since SiRGRaF has also been implemented on Solar Orbiter Metis data by \cite{Telloni_2022}, it can be used in the SITCoM as input along with the above mentioned data sets.  Along with these data sets, the application of SiRGraF on the extreme ultraviolet (EUV) dataset will require detailed analysis and testing including image processing to enhance the structures observed in EUV passbands. The future updates of SITCoM will be equipped with such image-enhancing features to carry out the analysis presented here.
Users have the capability to save the processed movie and individual frames in various formats. Moreover, SITCoM offers different intensity scaling, such as MinMax, Logarithmic, and Gamma, allowing users to enhance specific features based on their preferences and requirements.

To facilitate kinematics analysis of any sunward or anti-sunward moving structures in the data, SITCoM empowers users to generate height-time plots at different position angles. These plots provide valuable insights into the temporal evolution of the observed phenomena. The users can perform linear or higher-order fitting on these plots to derive essential parameters such as velocity or acceleration, depending on the type of fitting performed. Multiple measurements can be made at the same position angle or over desired angles to get an estimate of deviation in the measured kinematics throughout the large-scale structures.

Users can also investigate various types of oscillations and waves within a selected slit across a dynamic feature. By employing a decaying sine wave fitting technique, users can automatically or manually fit the resultant distance-time plot generated from the slit. The extracted parameters of the fitted wave can be saved for further analysis, providing insights into the oscillatory behavior of coronal features. 

This GUI is not only a powerful tool for analyzing coronal dynamics but also saves time typically spent on plotting and fitting. By streamlining the analysis process, SITCoM provides useful results obtained with quick analysis. The combination of advanced algorithms and user-friendly tools facilitates a comprehensive understanding of the observed phenomena, enhancing our knowledge in this field.
In the future, we plan to develop SITCoM for different platforms to make it more accessible to researchers worldwide. This will enable more scientists to benefit from its capabilities and contribute to our understanding of the Sun's corona.

We plan to improve the SITCoM application with the addition of new features in the future. These include the implementation of fitting multiple ridges identified in the height-time plots for kinematics analysis. As a result, multiple transients at the same position angle can be tracked and characterized manually without having to generate the same x-t map repeatedly.
Implementation of curved slit selection will enable a user to select a curved slit for the generation of x-t maps. Such a feature will be helpful to characterize the CME deflection, and non-radial flows \citep{Alzate2023ApJ} as well as tracking the waves and oscillations at bigger regions of interest using a single large slit or a wide slit. A provision to save the coordinates corresponding to the slit location and the x-t map array in the form of FITS or `.sav' file will be useful to revisit an analysis later or import to any other program when required.


These planned advancements will further enhance the capability of SITCoM and will be a part of its future releases.


\section*{Conflict of Interest Statement}

The authors declare that the research was conducted in the absence of any commercial or financial relationships that could be construed as a potential conflict of interest.

\section*{Author Contributions}

RP planned the development and overlooked the implementation of the tool. PU and AKS developed the tool in Python and carried out tests on different datasets. PU and AKS prepared the manuscript. All the authors took part in the discussion of the results.

 \section*{Funding}
A.K.S is supported by the Council of Scientific \& Industrial Research (CSIR), India, under file no. 09/079(2872)/2021-EMR-I. 

\section*{Acknowledgments}
PU and AKS would like to thank ARIES for providing the computational facilities. 
The SECCHI data used here were produced by an international consortium of the Naval Research Laboratory (USA), Lockheed Martin Solar and Astrophysics Lab (USA), NASA Goddard Space Flight Center (USA), Rutherford Appleton Laboratory (UK), University of Birmingham (UK), Max-Planck-Institut for Solar System Research (Germany), Centre Spatiale de Li$\grave{e}$ge (Belgium), Institut d'Optique Th$\acute{e}$orique et Appliqu$\acute{e}$e (France), Institut d'Astrophysique Spatiale (France). We also thank NASA for making SOHO/LASCO data publicly available.


 \section*{Data Availability Statement}
 The STEREO/SECCHI and SOHO/LASCO data sets analyzed for this study can be found in their respective data archives under the open data policy. The data sets generated in this study can be made available upon request.

\bibliographystyle{frontiersinSCNS_ENG_HUMS} 
\bibliography{main_v1}

\end{document}